\documentclass[fleqn,12pt,twoside]{article}
\usepackage{espcrc1}

\usepackage{graphicx}
\usepackage[figuresright]{rotating}

\usepackage{wrapfig}

\newcommand{\AmS}{{\protect\the\textfont2
  A\kern-.1667em\lower.5ex\hbox{M}\kern-.125emS}}

\newcommand{\ssnnm}{\sqrt{s_{\mbox{\tiny NN}}}}
\newcommand{\ssnn}{$\ssnnm$}
\newcommand{\snn}[1]{{\ssnn} = #1 GeV}

\newcommand{\etal}{$et$ $al.$}
\newcommand{\pt}{$p_T$}
\newcommand{\mt}{$m_T$}
\newcommand{\mpt}{$\langle${\pt}$\rangle$}
\newcommand{\pbar}{$\overline{p}$}
\newcommand{\coll}[1]{$#1$+$#1$}
\newcommand{\pp}{\coll{p}}
\newcommand{\ppp}{$p$(\pbar)+$p$}
\newcommand{\AuAu}{\coll{Au}}
\newcommand{\AAA}{\coll{A}}
\newcommand{\PbPb}{\coll{Pb}}
\newcommand{\ee}{$e^+$+$e^-$}
\newcommand{\OmegaBar}{$\overline{\Omega}^+$}

\title{Soft Physics in STAR}

\author{G. Van Buren\address[MCSD]{Brookhaven National Lab,
        Upton, New York 11973-5000, USA}
             for the STAR Collaboration\footnote{For the full author list
and acknowledgements, see Appendix ``Collaborations'' of this volume.}}
       
\begin{document}

\maketitle

\begin{abstract}
The STAR Experiment at RHIC is well-suited to making measurements of
particle yields from relativistic nuclear collisions
in the low transverse momentum, or ``soft'', regime. We present
preliminary results on measurements of
$\pi^0$, $\pi^-$, $\pi^+$, $K^-$, $K^+$, $K^0_s$, $\rho$,
$K^{*0}$+$\overline{K^{*0}}$, f$_0$, $p$, \pbar, $\phi$,
$\Lambda$, $\overline{\Lambda}$, $\Xi^-$, $\overline{\Xi}^+$, $\Omega^-$, and
\OmegaBar~from the first two years of physics
running at RHIC. An abundance
of physics topics can be addressed by these measurements,
some of which are discussed here along with the results.
\end{abstract}

\section{Introduction}

\renewcommand{\thefootnote}{\fnsymbol{footnote}}

``The principles of thermodynamics are devised for giving a phenomenological
account of the gross behavior of macroscopic physical systems,
...[interpreting] the average behavior of systems of many degrees
of freedom.''\footnotemark \footnotetext{R. Tolman, {\it The Principles of Statistical
Mechanics}, Oxford University Press (1938).}
The ongoing search for a system of nuclear matter which can be
described by a thermodynamic quark gluon plasma phase (QGP)
must therefore involve a study of nuclear matter in bulk.
For this reason, heavy ions with significant
amounts of nuclear matter are used in laboratory
experiments to find and study QGP. This also implies that
QGP formation affects some bulk portion of the outgoing matter,
most of which is in the ``soft'', low-\pt~region of phase space.

The STAR Experiment at RHIC has measured a rich variety of
particle species and portions of their soft spectra in high
energy \AuAu~and
\pp~collisions~\cite{Jen,Fuqiang,Javier,Christophe,Geno,Patricia,Ian,Falk}.
Because production, annihilation, and rescattering processes
differ among the particle species, their abundances and spectra
reflect on different aspects of the collision evolution. We present
here a discussion of these aspects as elucidated by our
measurements. We will follow an outline of the physics
topics along the evolution from initial collisions
to final reinteractions of the products.

\section{Inclusive Multiplicity and Mean Transverse Momenta}

\begin{wrapfigure}{r}{0.51 \textwidth}
\begin{center}
\vspace{-1.3cm}
\includegraphics[width=0.49 \textwidth]{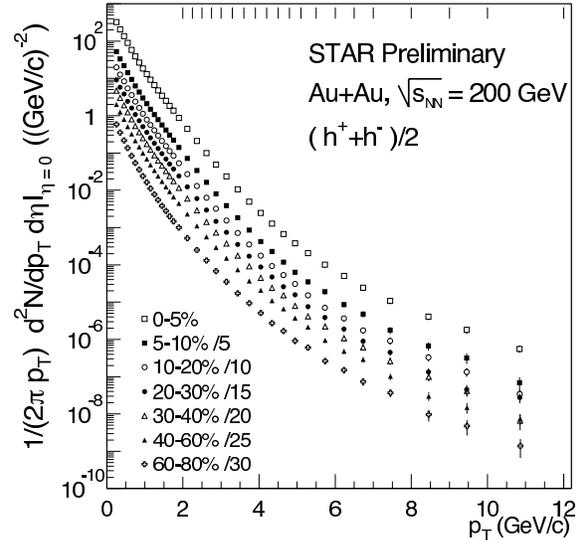}
\vspace{-0.9cm}
\end{center}
\caption{\footnotesize Transverse momentum spectra of
hadrons from \snn{200} \AuAu~over
several centralities. Only statistical errors are shown.
See Ref.~\cite{Jen} for systematic errors and more details.}
\vspace{-0.5cm}
\label{fi:spectra}
\end{wrapfigure}

Improved statistics in the second year of physics running at RHIC have
given STAR the ability to measure unidentified charged hadron transverse
momentum (\pt) spectra from 0.2 GeV/$c$ to
$\sim$12 GeV/$c$. These spectra are shown in Fig.~\ref{fi:spectra} as a
function of centrality for \AuAu~collisions.
They have been corrected for decays
and backgrounds through detailed simulation~\cite{hminus}. The corrections
are \pt~and multiplicity dependent, but nowhere exceed 20\%.
From these inclusive spectra we
can measure total yields as well as \mpt. This is done by fitting the spectra
with a power law of the form $C (1 + p_T/p_0)^{-n}$~\cite{highpt}. The
\pt-integrated multiplicity requires extrapolation of the power law at
the low-\pt~end, which comprises up
to 21\% of the yields. 
For the 5\% most central
events, our preliminary results indicate that
the average charged hadronic multiplicity
$\langle$$dN_{ch}/d\eta$$\rangle$
increases from 563$\pm$39 to 687$\pm$69
between \snn{130 and 200}.
\mpt~is then obtained from the same power law, giving
0.520$\pm$2.3\% GeV/$c$ at \snn{130} and
0.524$\pm$2.3\% GeV/$c$ at 200 GeV.
For both $\langle$$dN_{ch}/d\eta$$\rangle$
and \mpt~measurements the errors are dominated
by systematics mostly common between the two energies
and do not reflect possible relative variations between
each other.

\begin{wrapfigure}{r}{0.5 \textwidth}
\begin{center}
\vspace{-1.5cm}
\includegraphics[width=0.46 \textwidth]{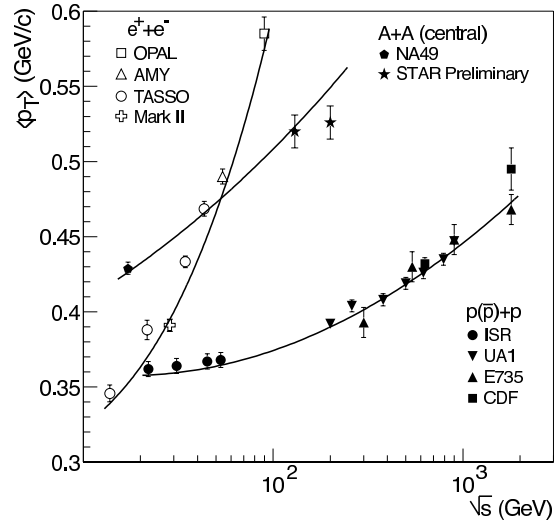}
\vspace{-0.8cm}
\end{center}
\caption{\footnotesize Mean transverse momenta of hadrons
for central heavy ion, \ee,
and \ppp~collisions versus energy~\cite{Xus}. Curves are CGC model
for \AAA~, phenomenological parameterization for \ppp,
and JETSET calculations (versus thrust axis) for \ee.
STAR data errors are systematics-dominated.}
\vspace{-2.3cm}
\label{fi:pt_scaling}
\end{wrapfigure}

To put these measurements in context,
a parameterization of the Color Glass Condensate (CGC)
model~\cite{GSM} is shown fit to central \AAA~\mpt~at
\snn{17 and 130} in Fig.~\ref{fi:pt_scaling}~\cite{Xus}.
While the CGC parameterization can fit these two data points,
the 200 GeV central \AuAu~\mpt~datum falls
well below the fit from the lower energies,
despite anticipation that the model is most applicable in
high energy central \AAA~collisions. The CGC model has
had numerous other successes~\cite{Dima}, but may
require supplemental considerations to congrue with
these data~\cite{Mueller}.

Also evident in Fig.~\ref{fi:pt_scaling} are the
differences in collision energy systematics for
\mpt~between \AAA, \ee~and \ppp~data.
This implies that \AAA~is not just a simple
superposition of nucleon+nucelon~collisions~\cite{Mueller},
and argues against a universality in particle
production~\cite{Ullrich}
suggested from total particle yields~\cite{Phobos}.

\section{Baryon Number}

Conserved quantities must survive any evolution. The conserved net-baryon
number is established from the incoming nuclei, but becomes
spread over the rapidity interval by initial interactions
(often called $stopping$, though not strictly stopped),
and then additionally smeared
by the rescattering processes; together these make up the
transport of baryon number from the nuclei to their final rapidities.
The midrapidity baryon and
antibaryon yields thus have folded into them these
transport distributions along with components from
baryon-antibaryon pair production (and annihilation, which
may also contribute to transport) processes.

\begin{wrapfigure}{r}{0.5 \textwidth}
\begin{center}
\vspace{-1.5cm}
\includegraphics[width=0.48 \textwidth]{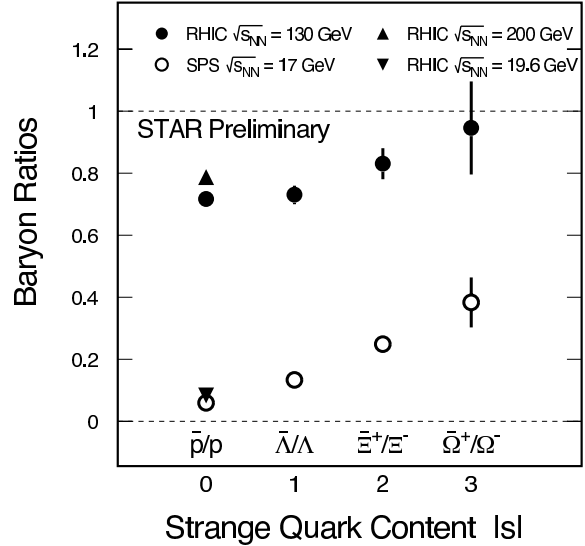}
\vspace{-0.7cm}
\end{center}
\caption{\footnotesize
Midrapidity antibaryon to baryon ratios are shown for
various species
and at several energies of central heavy ion collisions.
Only statistical errors are shown. Systematic errors range
from $\sim$7\% for \pbar/$p$ to $\sim$15\% for
$\overline{\Omega}^+/\Omega^-$ in RHIC data.}
\label{fi:baryons}
\vspace{-0.6cm}
\end{wrapfigure}

A characterization of the baryon-an\-ti\-bary\-on differences is made
in Fig.~\ref{fi:baryons}.
This plot shows the antibaryon to baryon ratios at midrapidity
for various species in central heavy ion collisions. It is clear from the
plot that the
ratios tend towards one with increasing strangeness content, and with
increasing energy. The ratios are not corrected for feeddown, and
the $p$ and \pbar~are essentially inclusive. These ratios indicate that
the yield of baryons resulting from pair production processes has
begun to dominate over those from transport
at RHIC energies:
the ratio of those contributions from inclusive (anti)protons in central \AuAu~is
$\sim$2.5 at \snn{130}, derived from inclusive
\pbar/$p$=0.72$\pm$0.05($syst.$),
and $\sim$3.5 at 200 GeV,
from \pbar/$p$=0.78$\pm$0.05($syst.$).
Inclusion of strange quarks in the baryons appears to dilute the contributions
of transport, enough to make the $\overline{\Omega}^+/\Omega^-$ ratio
consistent with 1 at RHIC; this is not the case at SPS energies where the
contributions from transport are still evident in the $\Omega$ yields
despite an initial net strange quark number of zero.

STAR has also measured most of these antibaryon to baryon ratios as a function
of centrality and sees only a slight rise in the antiproton
to proton ratio for peripheral collisions, and no significant
changes for the other ratios~\cite{Fuqiang,Javier,LambdaPaper}.
This implies little change in
the relevance of pair production processes for baryon and
antibaryon production despite the changing collision geometry
and participants.

\begin{figure}
\begin{center}
\vspace{-0.2cm}
\includegraphics[width=0.50 \textwidth]{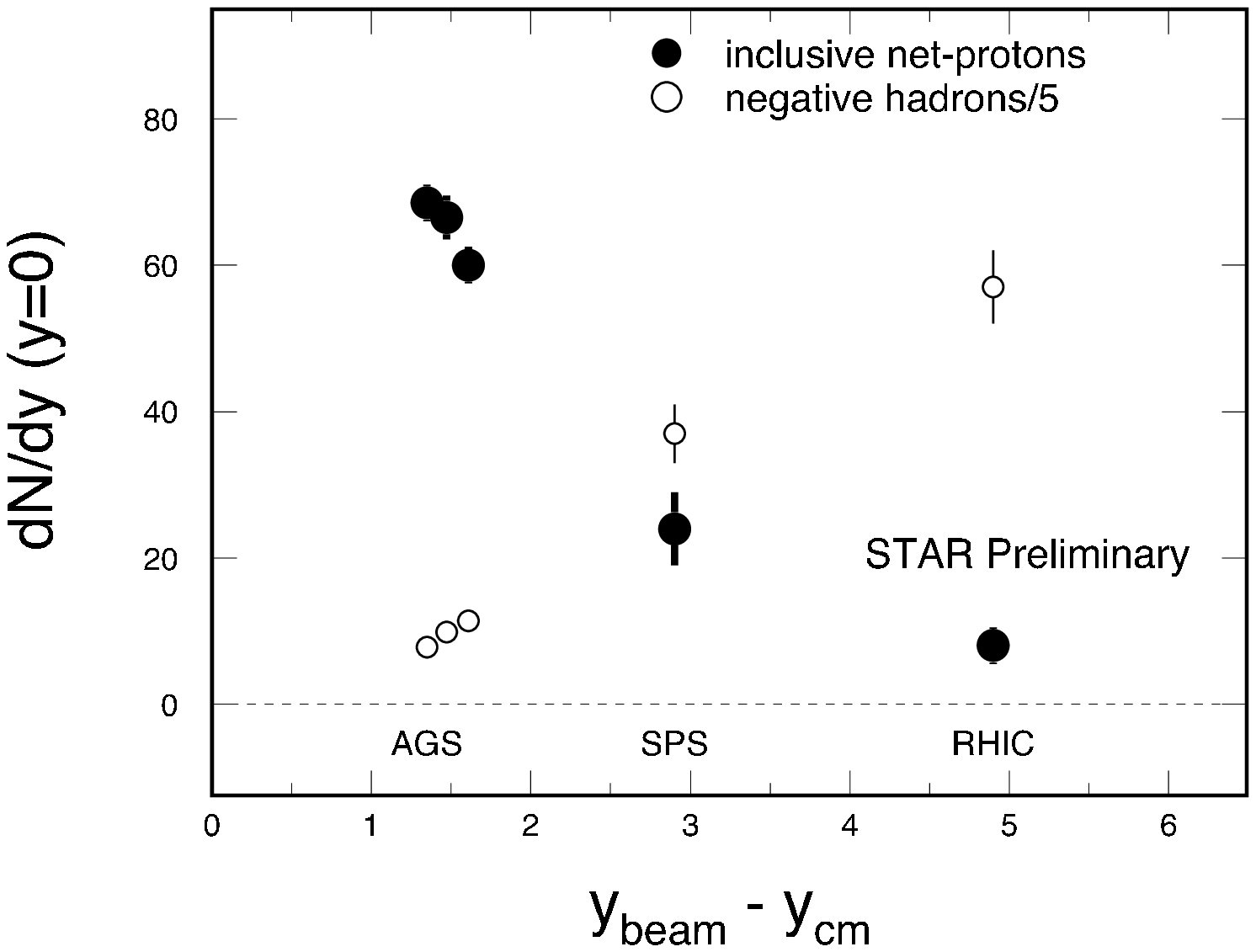}
\hfill
\includegraphics[width=0.38 \textwidth]{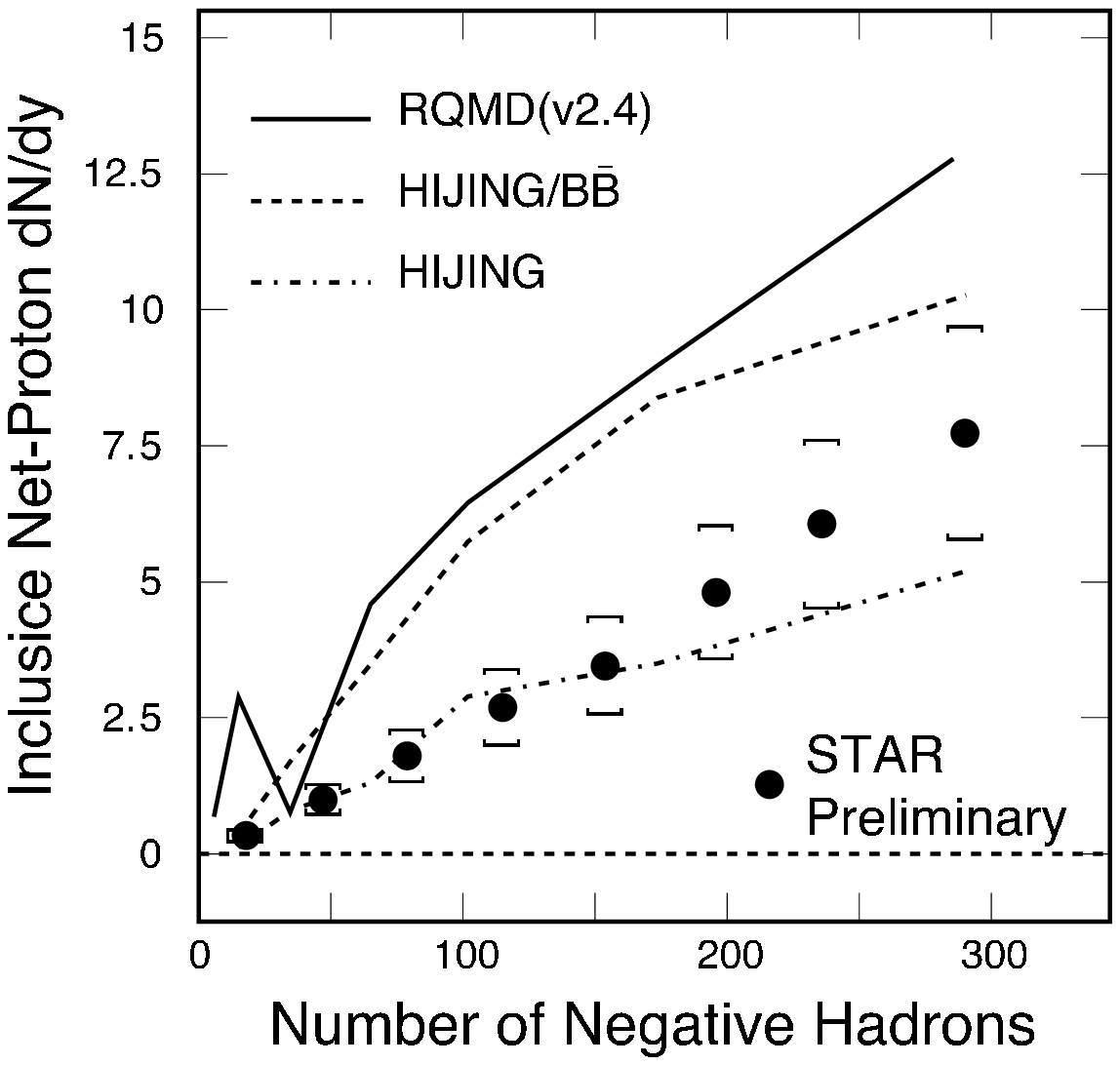}
\vspace{-1.2cm}
\end{center}
\caption{\footnotesize
The net
protons at midrapidity are shown versus collision energy
for central collisions along with
the negative hadrons (left, statistical errors only),
and versus centrality for \snn{130} along with
model predictions (right, errors systematics-dominated).}
\label{fi:netp}
\vspace{-0.8cm}
\end{figure}

A definitive measure of the baryon number transported to
midrapidity comes from the difference between yields of baryons
and antibaryons at midrapidity. This is represented by the net inclusive
proton number shown in Fig.~\ref{fi:netp}(left). This number
is clearly decreasing with increasing collision energy for large
heavy ion collisions, while the yield of produced particles at midrapidity
(represented in Fig.~\ref{fi:netp}(left) by the negative hadron
yield) is monotonically increasing. This demonstrates the decreasing
significance of baryon transport contributions to midrapidity
yields and is not unexpected as the rapidity interval from the
original nuclei is stretched. It is interesting that the
baryon number transport is in fact still nonzero despite the
interval of $\sim$5 units in rapidity at 130 GeV (also true for
$\sim$6 rapidity units at 200 GeV). Some understanding of this
may come from the asymmetric $q$ versus $\overline{q}$ distributions
even at low momentum fraction $x$=$p_{q,\overline{q}}/p_{N}$
within the incoming nucleons~\cite{Berndt}. Several model predictions
are also shown for the centrality plot, with HIJING matching peripheral
data best, and HIJING/B$\overline{\rm B}$ (which is intended
to enhance the stopping~\cite{HIJINGBB}) more closely
approaching the central data. RQMD overpredicts the transport
contributions.

\begin{figure}
\begin{minipage}{0.03 \textwidth}
\rotatebox{90}{All plots:~~~1/(2$\pi$\mt~N$_{\rm events}$)~($d^2$N/$d$\mt $d$y)~
~($c^4$/GeV$^2$)}
\end{minipage}
\begin{minipage}{0.97 \textwidth}
\begin{center}
\vspace{-0.2cm}
\includegraphics[height=7.0cm]{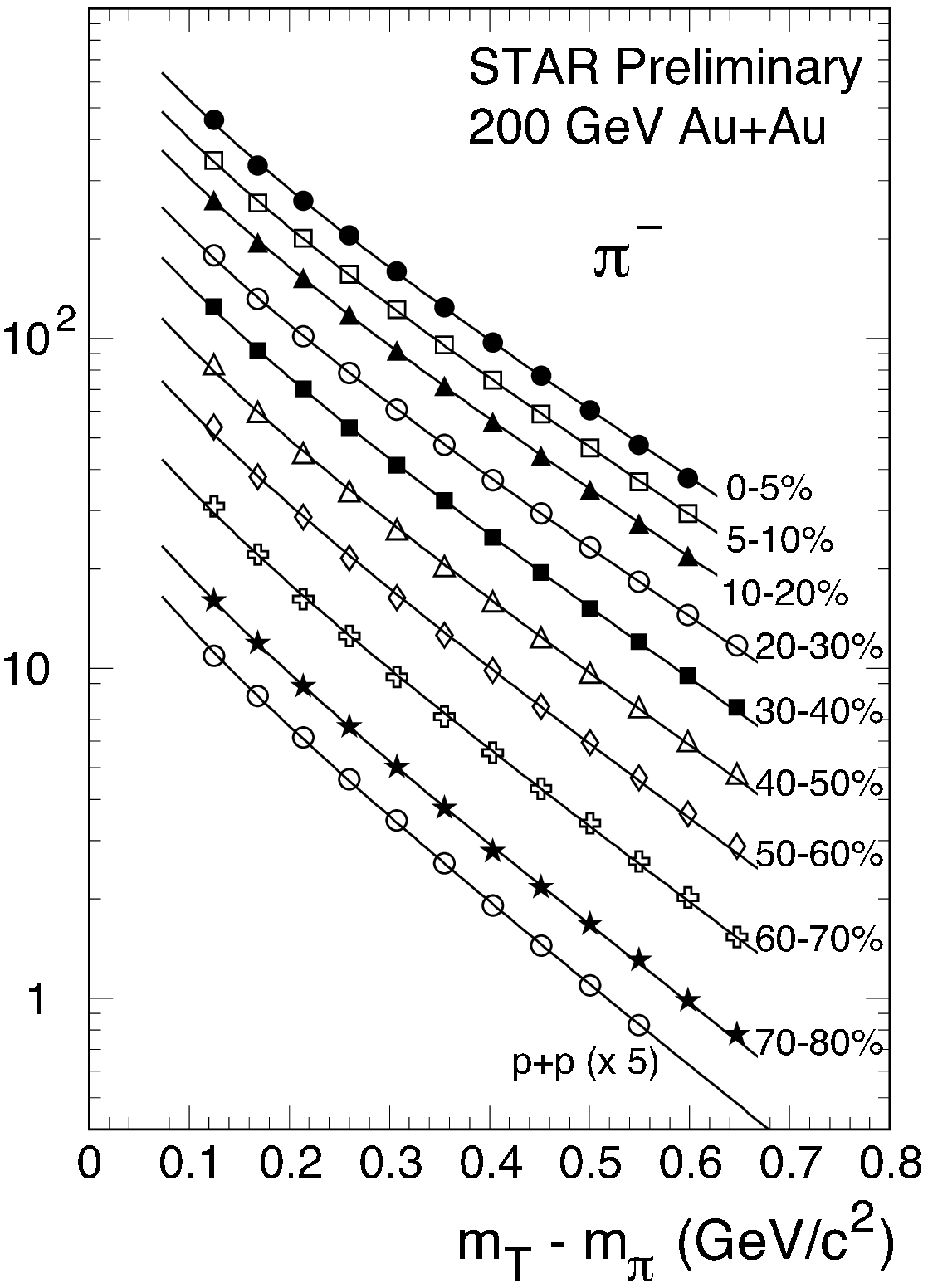}
\hfill
\includegraphics[height=7.0cm]{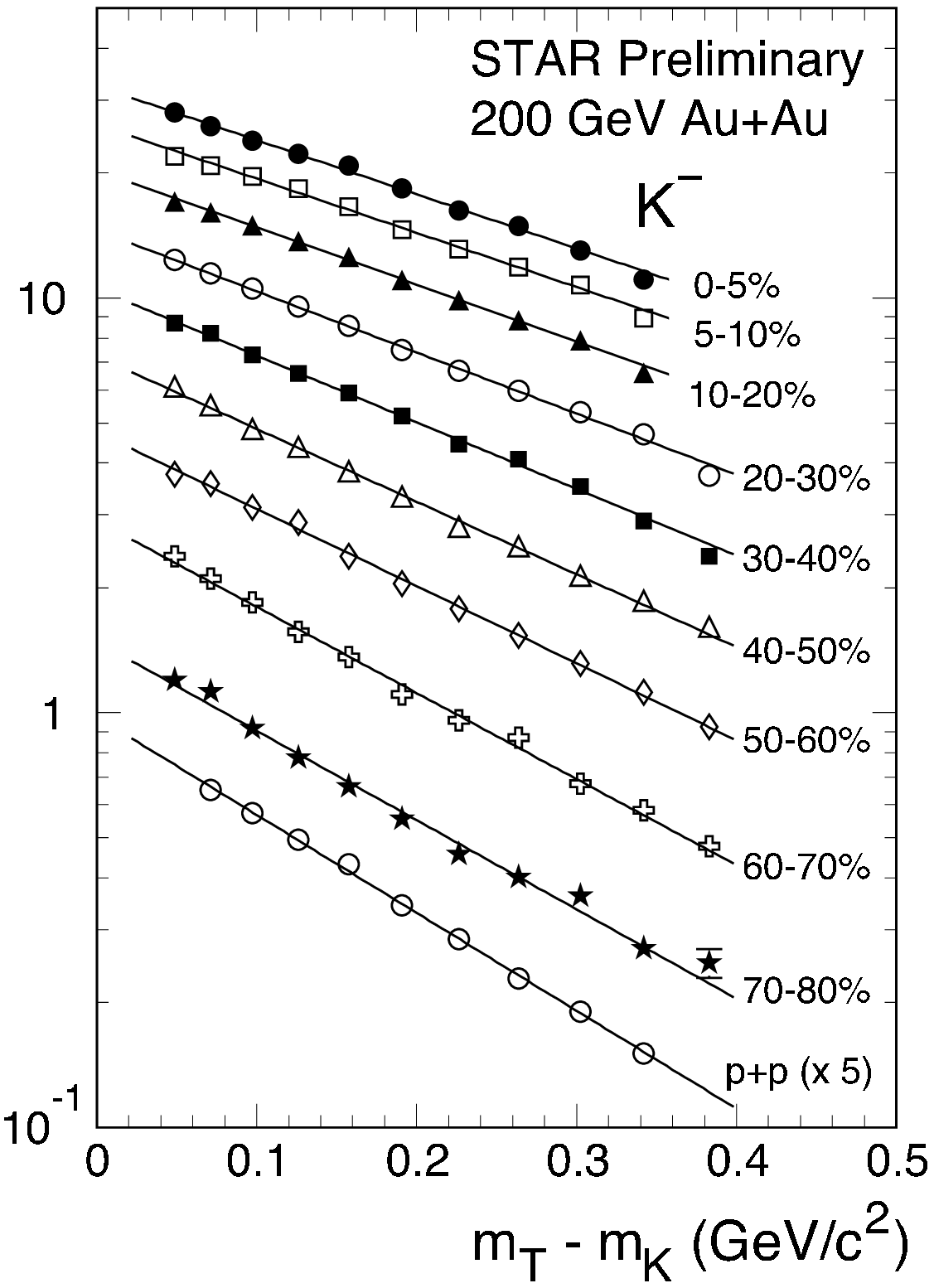}
\hfill
\includegraphics[height=7.0cm]{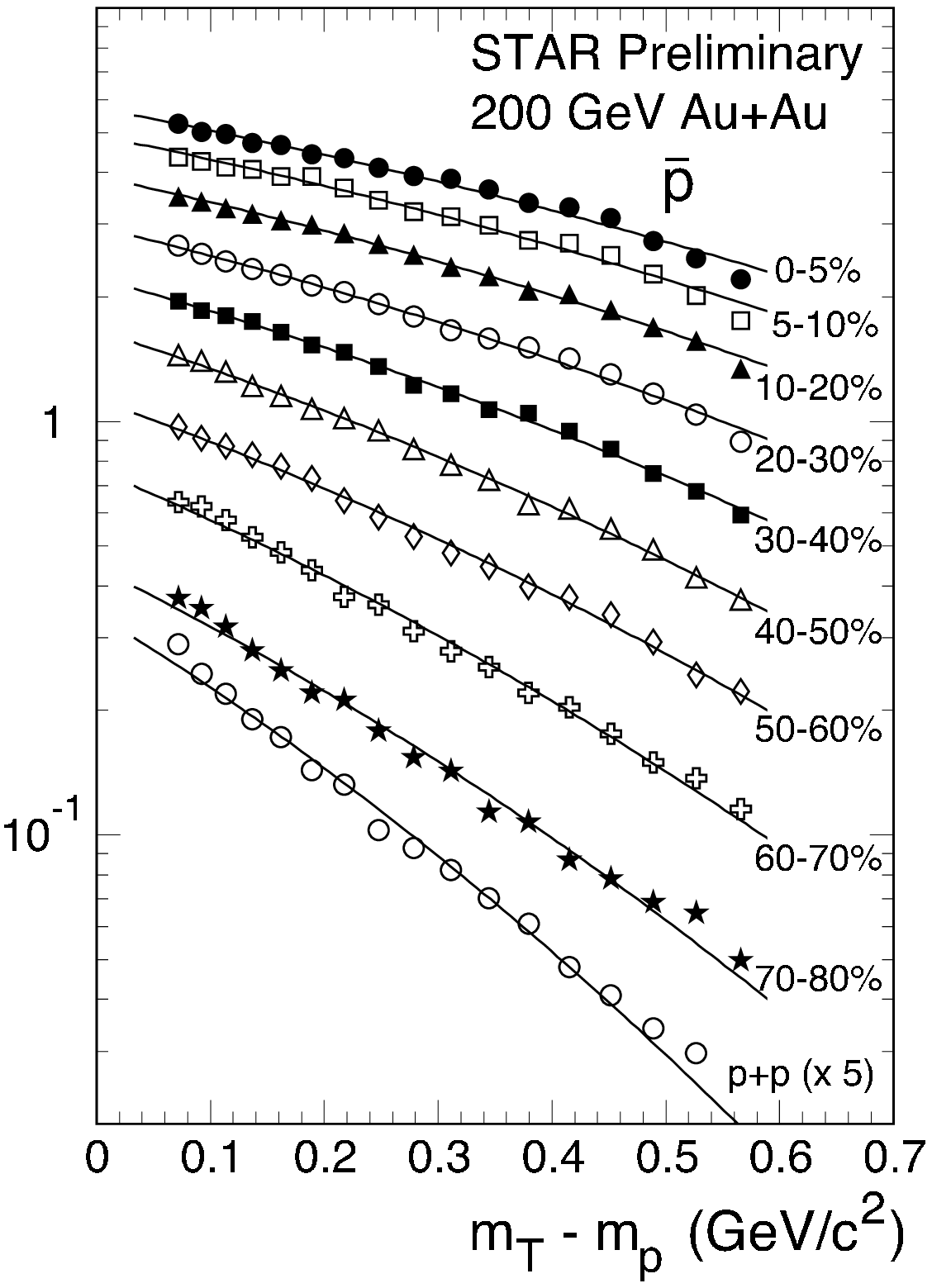}
\newline
\includegraphics[width=0.512 \textwidth]{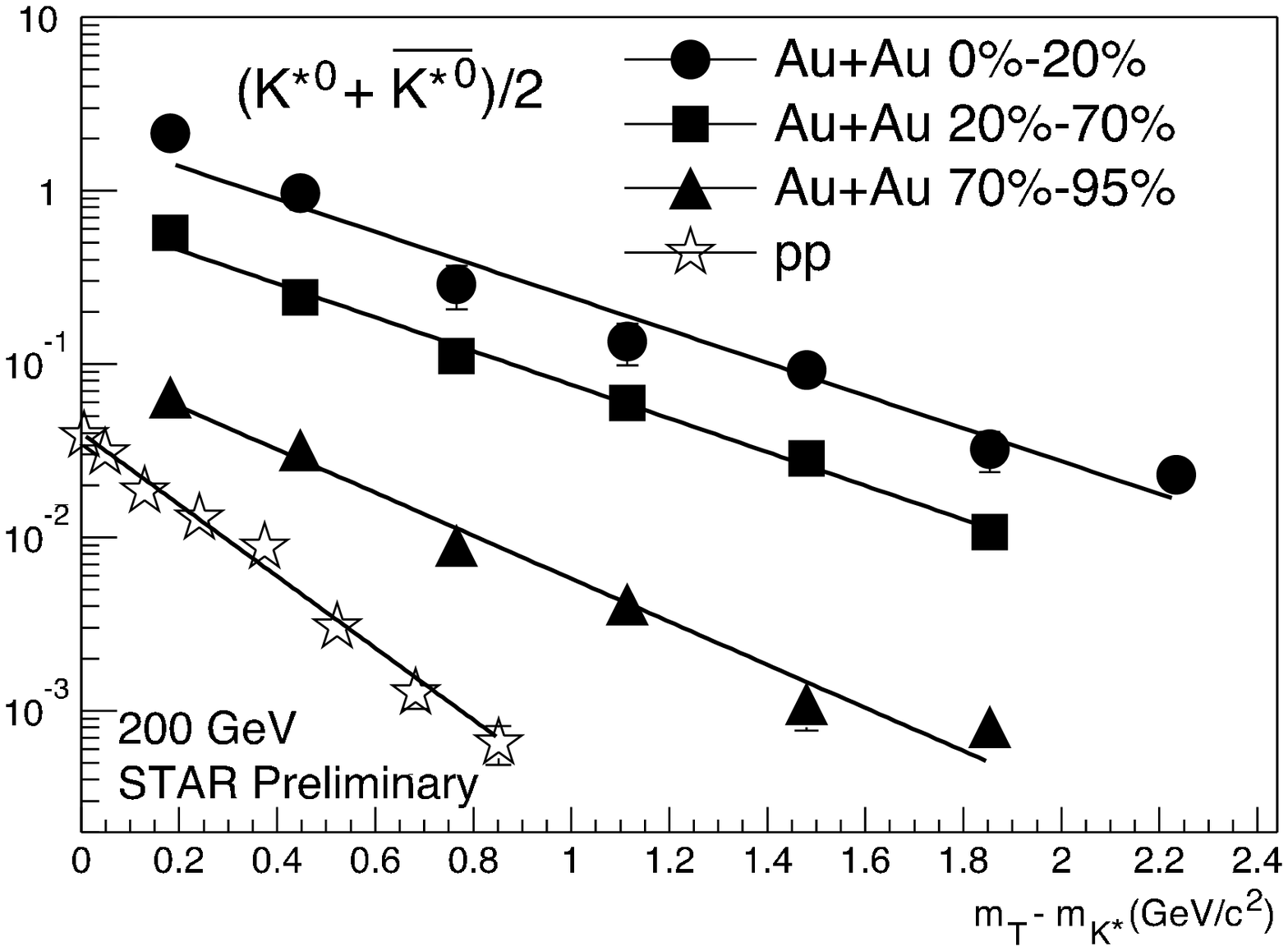}
\hspace{-0.7cm}
\includegraphics[width=0.517 \textwidth]{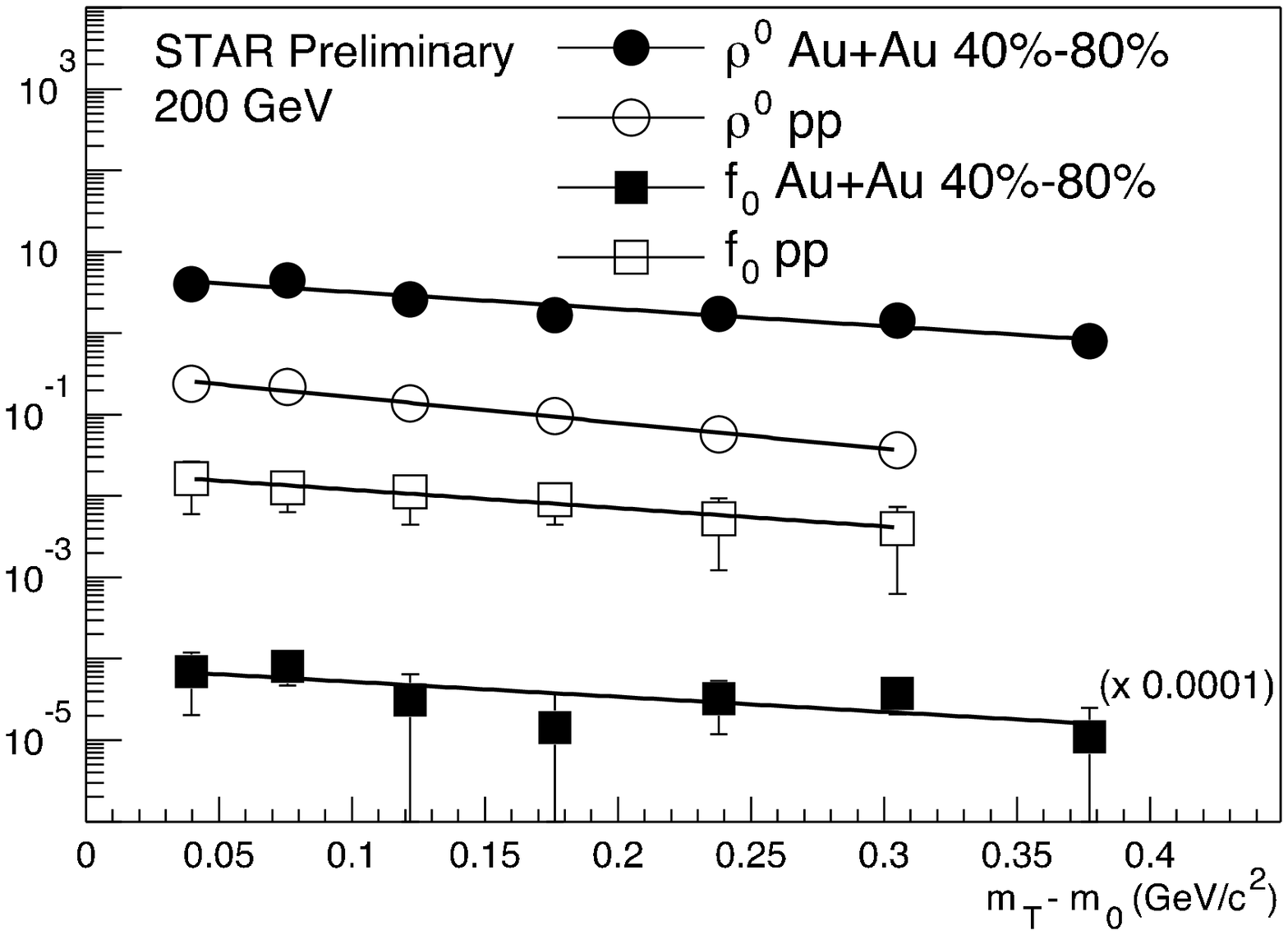}
\newline
\includegraphics[width=0.520 \textwidth]{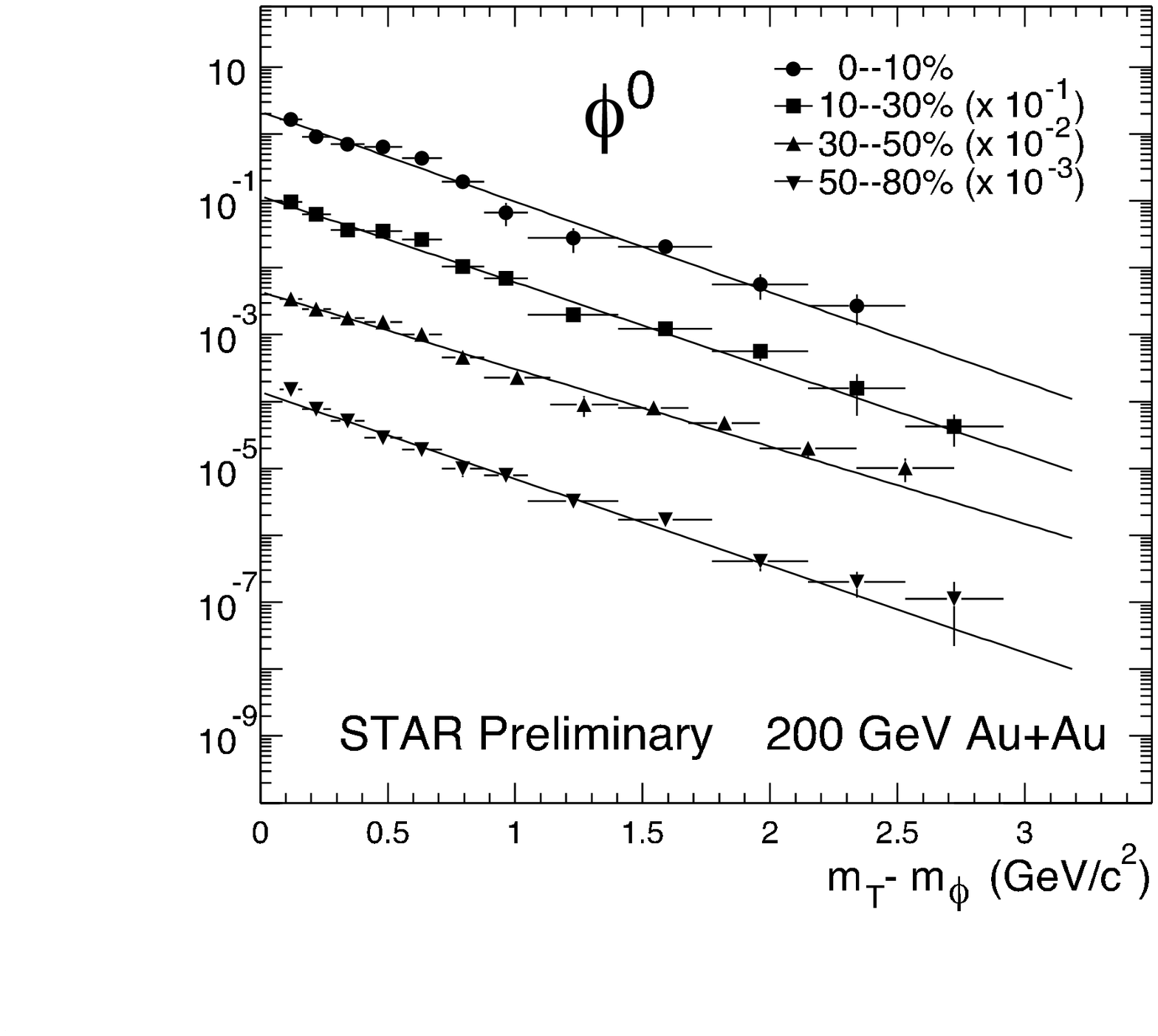}
\hspace{-0.6cm}
\vspace{-1.3cm}
\includegraphics[width=0.5 \textwidth]{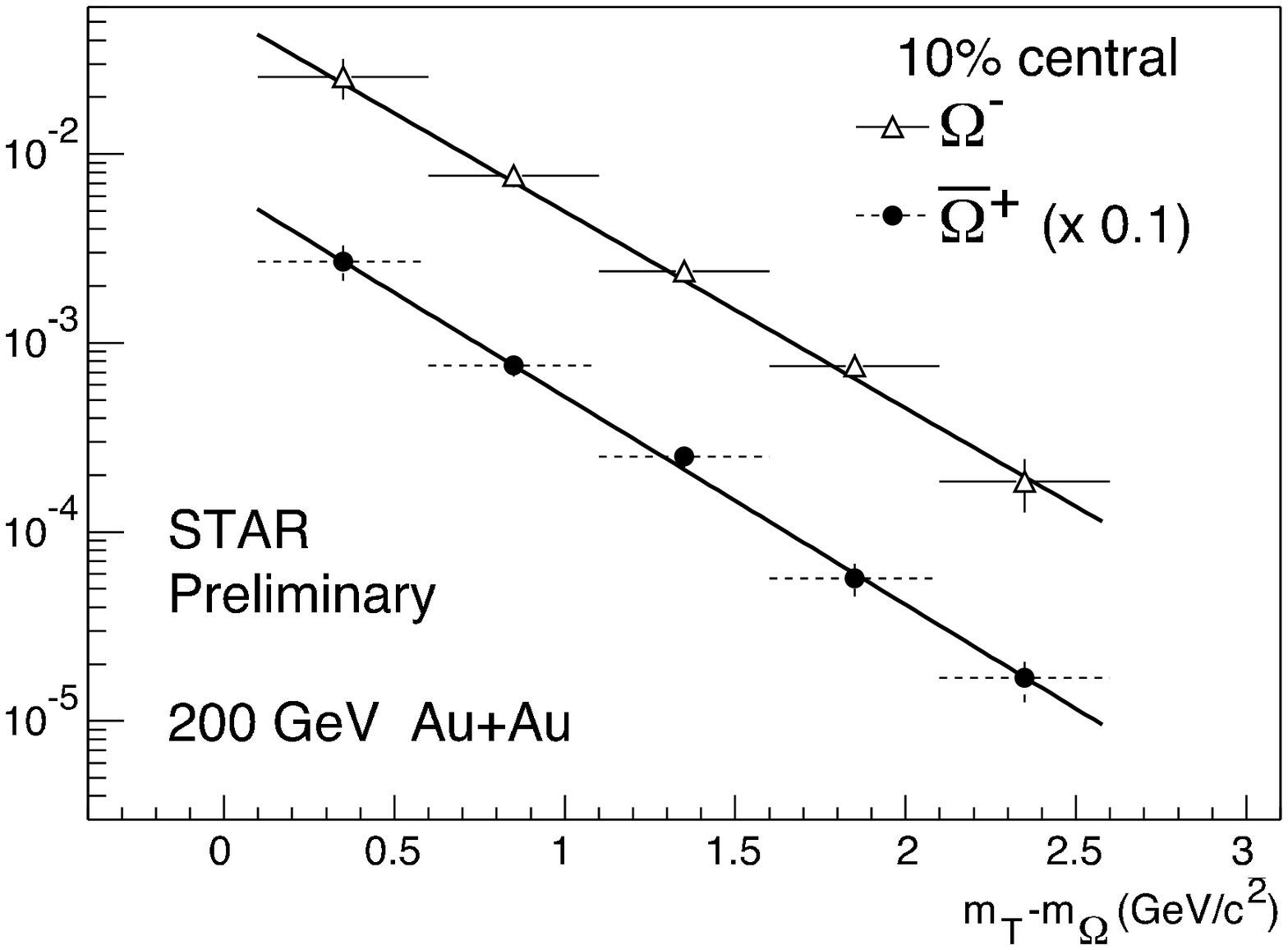}
\end{center}
\end{minipage}
\caption{\footnotesize A variety of preliminary midrapidity
spectra from \snn{200}
\AuAu~and \pp~collisions in STAR are plotted,
including those of $\pi^-$,
$K^-$, and inclusive \pbar~\cite{Fuqiang},
($K^{*0}$+$\overline{K^{*0}}$)/2, $\rho$, and f$_0$~\cite{Patricia},
$\phi$~\cite{Geno}, and $\Omega$~\cite{Christophe}.
Only statistical errors are shown. See respective references for
systematic errors and further details.
All fits are \mt~exponentials except for Bose-Einstein on $\pi^-$ and
\pt~Gaussian
on \pbar; the fit results can be found in Table~\ref{ta:spectra}
for selected centralities.}
\label{fi:AllSpectra}
\end{figure}

\section{Particle Production}

Long-standing arguments exist for extraordinary hadron
yields should a QGP form, such as strangeness phase space
saturation (enhanced yields over a solely hadronic
evolution)~\cite{QGPyields}
and J/$\psi$ suppression~\cite{jpsi_supp}.
While studies of hadron production in
the low \pt~region of phase space where a QGP is
likely to manifest are appropriate for these searches,
it should be understood that experiments often measure yields after
any QGP has condensed, at the end of all hadrochemical reactions
which can potentially hide earlier irregularities (equilibration of
a hadron gas, for example).

It is all the more important that we study the particle
production to learn what we can about the system
and its evolution.
STAR is well-suited for this task, capable
of identifying and measuring spectra and yields for a plethora
of particle species. Results on the following
species have been presented at this conference:
$\pi^0$, $\pi^-$, $\pi^+$, $K^-$, $K^+$, $K^0_s$, $\rho$,
$K^{*0}$+$\overline{K^{*0}}$, f$_0$, $p$, \pbar, $\phi$,
$\Lambda$, $\overline{\Lambda}$, $\Xi^-$, $\overline{\Xi}^+$, $\Omega^-$, and
\OmegaBar~\cite{Fuqiang,Javier,Christophe,Geno,Patricia,Ian,Falk}.
The impressive preliminary
spectra and fits for several of these species from the \snn{200}
data can be seen in
Fig.~\ref{fi:AllSpectra}, and in Table~\ref{ta:spectra}
where the \mt~reach of the measurements and results
of the fits are presented. Our observations of the $K^{*0}$, $\rho$,
and f$_0$ are the first in heavy ion collisions.

One can attempt to obtain a global characterization of
particle production by comparing the yields
to a model which describes the bulk properties of the
system through a limited set of parameters.
A thermal, chemical equilibrium model is one such statistical model,
and we can examine how this model describes
our \snn{130} midrapidity data, 
mindful of any caveats~\cite{thermal_limitations}.
In the particular model we choose,
free parameters include the temperature ($T$), light
quark chemical potential ($\mu_q$), strange quark chemical
potential ($\mu_s$), and strangeness saturation factor
($\gamma_s$, where values less than one indicate incomplete
saturation)~\cite{stat_model}.
Shown in Fig.~\ref{fi:thermal_fit} is a fit
excluding the $K^*$,
$\phi$, $\Xi$, and $\Omega$ results ($\chi^2$/dof = 1.53).
From this we can compare the excluded data
to the outcome of the fit based on the included
species, demonstrating how well the model
matches all of the data.
Including all ratios results in $\chi^2$/dof = 1.68.
In either fit, the preferred value of $\gamma_s$ is close to 1.
Other chemical equilibrium models have also predicted and fit most
of these particle yields rather
well~\cite{other_fits}, converging on similar values
for the thermodynamic parameters (with a temperature range
from 165$\pm$7 MeV to 179$\pm$9 MeV, for example).
While these models may or may not be applicable,
and good fits are neither new
(even seen in \ee~and \ppp~data, albeit without strangeness saturation)
nor absolute proof of equilibrium~\cite{vkoch}, our results are provocative:
\begin{wrapfigure}{r}{0.60 \textwidth}
\begin{center}
\vspace{-1.25cm}
\includegraphics[width=0.59 \textwidth]{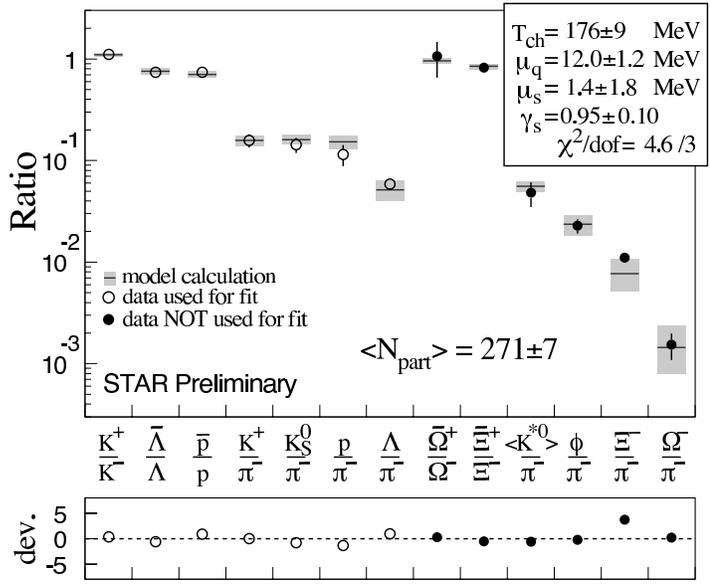}
\vspace{-0.6cm}
\end{center}
\caption{\footnotesize A fit to the STAR midrapidity 15\% central
130 GeV \AuAu~data with a thermal statistical model~\cite{stat_model}
is shown.}
\label{fi:thermal_fit}
\vspace{-0.8cm}
\end{wrapfigure}
1) the fits indicate approximate strangeness phase space
saturation for central collisions, and
2) recent lattice QCD
calculations for non-zero chemical potentials have determined a 
similar temperature of
172$\pm$3 MeV in this region of the phase boundary
between hadronic matter and QGP~\cite{lQCD}.

Table~\ref{ta:spectra} shows the preliminary results of spectra
measurements in STAR from \AuAu~at \snn{200} for $\pi^-$, $K^-$, $\rho$,
$K^{*0}$+$\overline{K^{*0}}$, \pbar, f$_0$, $\phi$,
$\Omega^-$, and $\overline{\Omega}^+$~\cite{Fuqiang,Christophe,Geno,Patricia}.
A preliminary \pt-independent muon contamination correction
has been applied to the pions.
More data is necessary to tightly constrain a chemical fit.

\begin{sidewaystable}[ht]
\begin{center}
\begin{tabular}{|c||*{6}{c|}}
\hline
Species & Centrality & \mt-$m_0$~(GeV/$c^2$) &
{\footnotesize Extrapolated} $dN/dy$ &
$T$ (MeV) & \mpt~(MeV/$c$) & $\chi^2$/dof \\
\hline \hline

& 0-5\% & &
353$\pm$15\% & 218$\pm$10\% & 408$\pm$10\% & 95/9 \\
\cline{2-2} \cline{4-7}
$\pi^-$ & 5-10\% & 0.10-0.62 &
270$\pm$15\% & 219$\pm$10\% & 410$\pm$10\% & 32/9 \\
\cline{2-2} \cline{4-7}
 & 10-20\%  &&
204$\pm$15\% & 216$\pm$10\% & 406$\pm$10\% & 53/9 \\
\cline{2-7}
 & 70-80\%  & 0.10-0.67 &
11.2$\pm$15\% & 185$\pm$10\% & 357$\pm$10\% & 14/10 \\
\hline \hline

  & 0-5\% &&
55.1$\pm$0.7$\pm$15\% & 327$\pm$5$\pm$10\% & 771$\pm$10$\pm$10\% & 36/9 \\
\cline{2-2} \cline{4-7}
$K^-$ & 5-10\% &  0.04-0.40 &
44.4$\pm$0.6$\pm$15\% & 326$\pm$5$\pm$10\% & 769$\pm$10$\pm$10\% & 44/9 \\
\cline{2-2} \cline{4-7}
 & 10-20\% &&
32.1$\pm$0.3$\pm$15\% & 310$\pm$3$\pm$10\% & 738$\pm$6$\pm$10\% & 43/9 \\
\cline{2-2} \cline{4-7}
 & 70-80\% &&
1.31$\pm$0.02$\pm$15\% & 202$\pm$5$\pm$10\% & 531$\pm$10$\pm$10\% & 16/9 \\
\hline \hline

$\rho$ &40-80\% &0.03-0.41 &
 6.59$\pm$1.08$\pm$30\% &207$\pm$54$\pm$30\% & 615$\pm$110$\pm$30\% &
1.63/5  \\
\hline \hline

& 0-20\% & 0.08-2.43 &
 8.35$\pm$0.99$\pm$25\% & 459$\pm$32$\pm$10\% & 1141$\pm$61$\pm$10\% &
11.2/5 \\
\cline{2-7}
($K^{*0}+\overline{K^{*0}}$)/2 & 20-70\% & 0.08-2.04 &
 2.68$\pm$0.18$\pm$25\% &447$\pm$20$\pm$10\% & 1118$\pm$38$\pm$10\% &
3.76/4 \\
\cline{2-7}
 & 70-95\% & 0.08-2.04 &
 0.28$\pm$0.03$\pm$25\% & 350$\pm$23$\pm$10\% & 931$\pm$45$\pm$10\% &
6.35/4 \\
\hline \hline

& 0-5\% &&
28.7$\pm$0.3$\pm$25\% & 894$\pm$6$\pm$10\%~\ddag & 1120$\pm$8$\pm$10\% & 71/15 \\
\cline{2-2} \cline{4-7}
\pbar & 5-10\% &  0.06-0.58 &
22.9$\pm$0.2$\pm$25\% & 861$\pm$6$\pm$10\%~\ddag & 1079$\pm$8$\pm$10\% & 81/15 \\
\cline{2-2} \cline{4-7}
 & 10-20\% & &
16.9$\pm$0.1$\pm$20\% & 830$\pm$4$\pm$10\%~\ddag & 1040$\pm$5$\pm$10\% &49/15 \\
\cline{2-2} \cline{4-7}
 & 70-80\% & &
0.84$\pm$0.01$\pm$15\% & 550$\pm$5$\pm$10\%~\ddag & 689$\pm$6$\pm$10\% & 38/15 \\
\hline \hline

f$_0$ & 40-80\% &0.02-0.35 & 
1.73$\pm$0.96$\pm$50\% &
200$\pm$177$\pm$50\% & 652$^{+355}_{-460}$$\pm$50\% & 1.37/5 \\
\hline \hline

 & 0-10\% &0.08-2.53 &
5.9$\pm$0.5$^{+1.0}_{-0.4}$ &
322$\pm$20$^{+25}_{-10}$ & 909$\pm$39$^{+49}_{-20}$ & 15.6/9 \\
\cline{2-7}
$\phi$ & 10-30\% & 0.08-2.91 &
3.4$\pm$0.2$^{+0.5}_{-0.3}$ &
337$\pm$16$^{+20}_{-20}$ & 939$\pm$32$^{+39}_{-39}$ & 24.4/10 \\
\cline{2-7}
& 30-50\% &0.08-2.72 &
1.4$\pm$0.1$^{+0.2}_{-0.1}$ &
375$\pm$16$^{+10}_{-50}$ & 1014$\pm$31$^{+20}_{-98}$ &  20.5/10 \\
\cline{2-7}
 & 50-80\% &0.08-2.91 &
0.40$\pm$0.02$^{+0.08}_{-0.02}$ &
334$\pm$15$^{+10}_{-40}$ & 953$\pm$30$^{+20}_{-79}$ & 12.5/10 \\
\hline \hline

$\Omega^-$ &0-10\% &0.10-2.60 &
0.31$\pm$0.04$\pm$15\% &417$\pm$23$\pm$15\% &
1269$\pm$47$\pm$15\% & 0.4/3 \\
\hline \hline

$\overline{\Omega}^+$ &0-10\% & 0.10-2.60 &
0.33$\pm$0.05$\pm$15\% &394$\pm$19$\pm$15\% &
1222$\pm$39$\pm$15\% &  1.2/3 \\
\hline
\end{tabular}
\vspace{-0.43cm}
\rotatebox{270}
{\Large \hspace{-5.5cm} STAR Preliminary \snn{200} \AuAu}
\end{center}
\caption{\footnotesize Preliminary results on midrapidity measurements
from \snn{200} \AuAu~collisions in STAR. The most central and most peripheral
measurements available are shown.
Inverse slope $T$, \mpt, and $\chi^2$/dof are from \mt~exponential fits,
except for Bose-Einstein on $\pi^-$ and \pt~Gaussian
on \pbar~(\ddag~where $T$ is replaced by the
Gaussian width $\sigma$).
First error quoted is statistical, second is systematic; if only one error is quoted,
the error is dominated by systematics. $\chi^2$/dof is large for $\pi^-$, $K^-$,
and \pbar~due to point-to-point systematics which are much larger than the
statistical errors.}
\label{ta:spectra}
\vspace{-1.0cm}
\end{sidewaystable}

\section{Transverse Momentum Systematics}

\begin{wrapfigure}{r}{0.465 \textwidth}
\begin{center}
\vspace{-3.3cm}
\includegraphics[width=0.444 \textwidth]{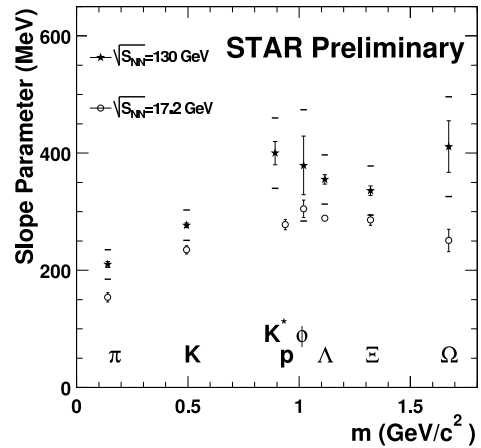}
\vspace{-0.55cm}
\end{center}
\caption{\footnotesize Inverse slope parameters as a function of particle
mass for SPS~\cite{sps_slopes,NuFlow} and RHIC are shown with
statistical error bars, and systematic error brackets for RHIC
data only.}
\label{fi:slopes}
\vspace{-1.05cm}
\end{wrapfigure}

\begin{figure}[t]
\begin{center}
\includegraphics[width=0.37 \textwidth]{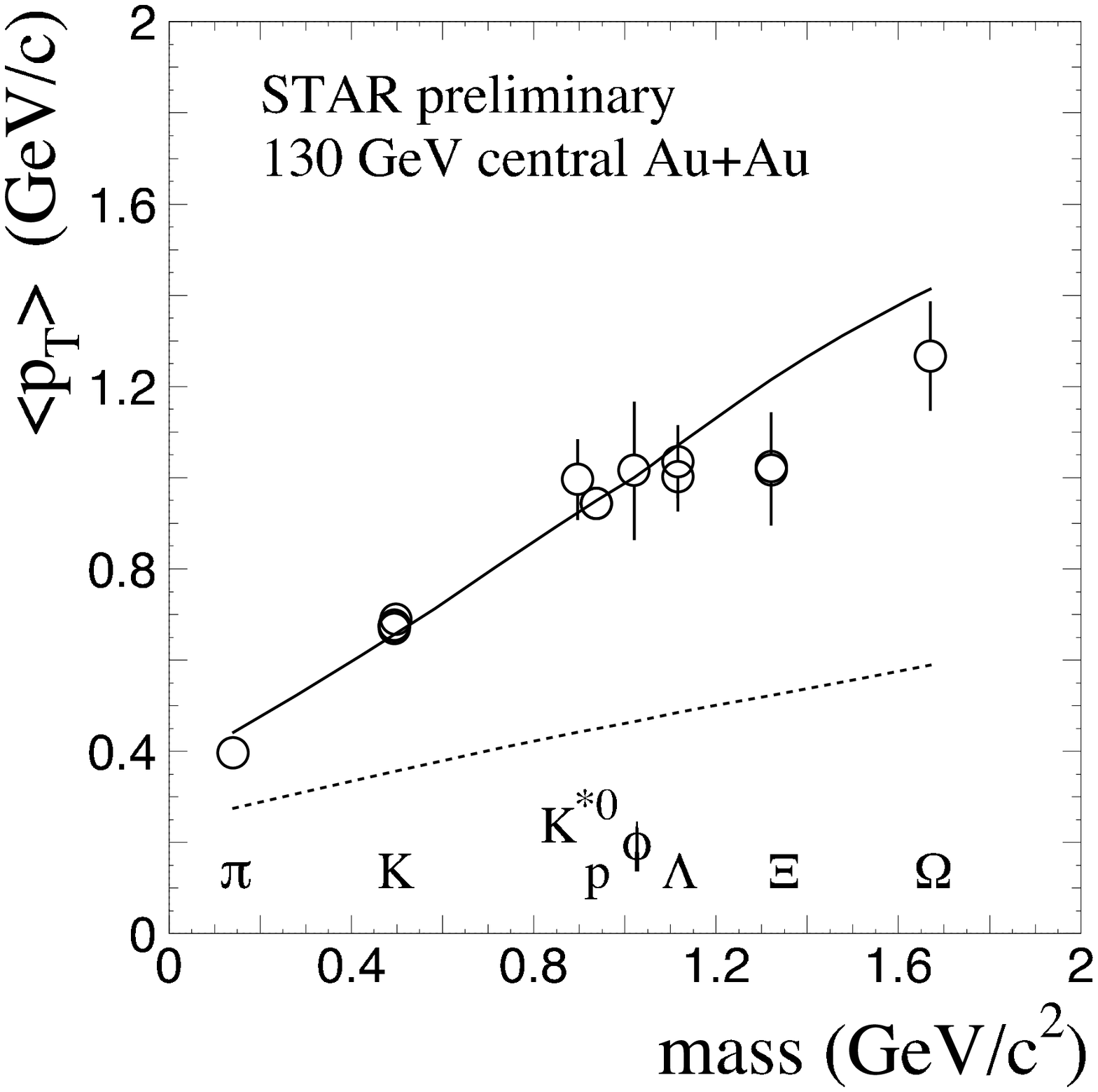}
\hfill
\includegraphics[width=0.5 \textwidth]{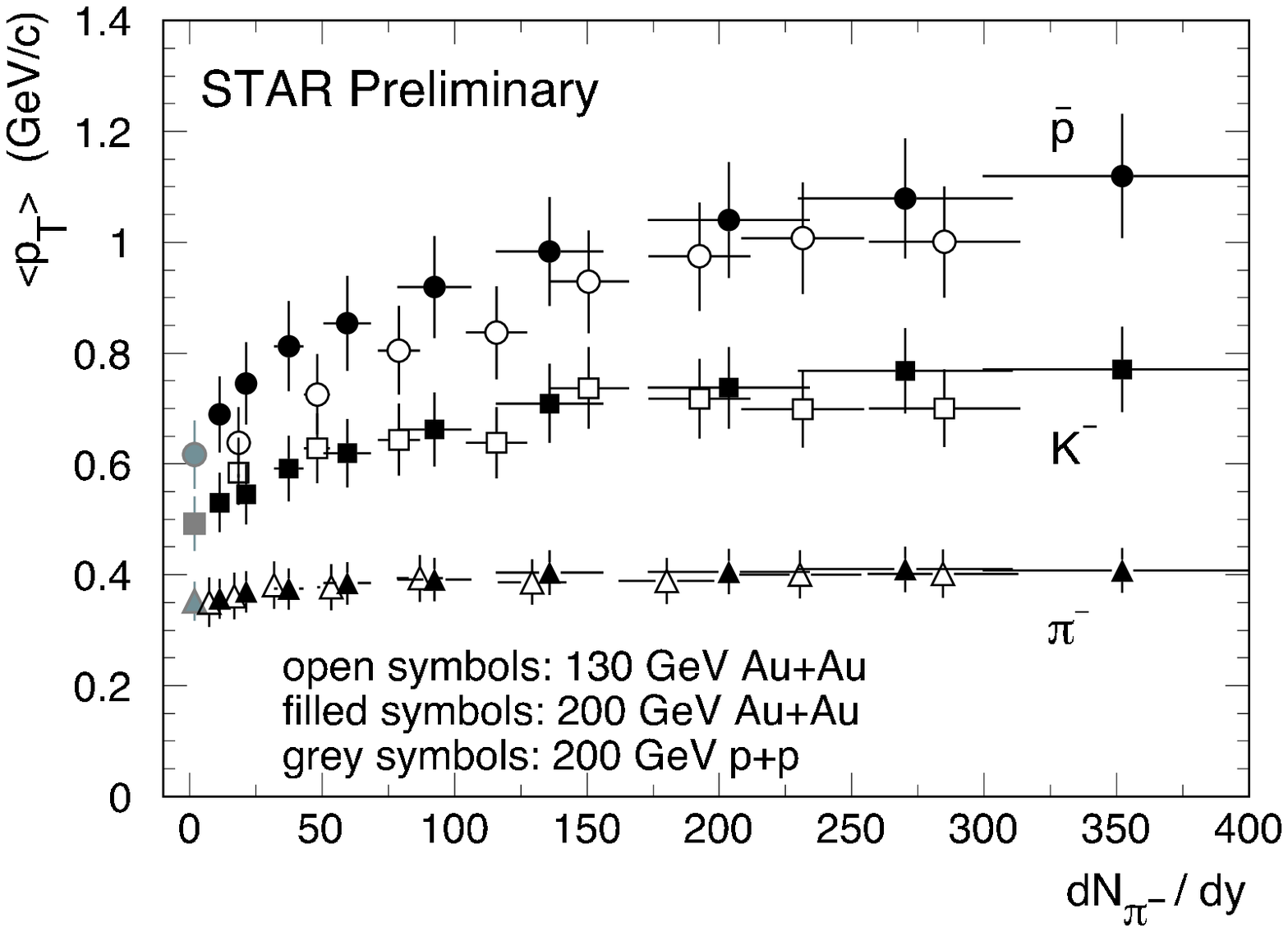}
\vspace{-1.1cm}
\end{center}
\caption{\footnotesize Mean transverse momenta of
identified species versus mass (left) for \snn{130} \AuAu~collisions
along with blast wave model fit (upper curve:
$T_{\rm fo}$$\simeq$100 MeV, $\langle\beta_{T}\rangle$$\simeq$0.55)
and zero flow curve (lower curve: $T_{\rm fo}$$\simeq$100 MeV, $\beta_T$$\equiv$0).
\mpt~of $\pi^-$, $K^-$, and \pbar~are mapped against 
multiplicity (right) for \snn{130} \AuAu, and 200 GeV \AuAu~and \pp.
All errors are systematics-dominated.}
\label{fi:meanpt}
\vspace{-0.6cm}
\end{figure}

Transverse radial flow can be gauged through the mass
dependence of either
inverse slope parameters ($T$)
of spectra fits (inversely proportional to the steepness of
yield fall-off in \mt)~\cite{NuFlow}, or \mpt~\cite{ptFlow}
of identified spectra.
These are shown for various STAR data from
central \AuAu~collisions at 130 GeV in Figs.~\ref{fi:slopes} and~\ref{fi:meanpt}(left).
It is important that the inverse slope is measured over a large
range of the produced spectrum to permit valid comparisons,
and this holds for the data shown in Fig.~\ref{fi:slopes}. Here it can be
seen that the rising trend versus mass expected from
transverse radial flow~\cite{NuFlow} is seen for the lighter species, but that
the strange baryons do not clearly follow this trend. Further discussion
of these systematics can be found in Ref.~\cite{Xus}.

The same behavior is exhibited in Fig.~\ref{fi:meanpt}(left)
by the central \mpt~values.
Fig.~\ref{fi:meanpt}(right) shows how the \mpt~develops
as a function
of centrality for $\pi^-$, $K^-$, and \pbar~at RHIC.
Going from \pp~data, through peripheral to central
\AuAu~data shows a smooth rise consistent with
the onset and buildup of transverse radial flow.

A blast wave~\cite{blast} fit to the central
$\pi^-$, $K^-$, and \pbar~spectra provides a transverse radial flow velocity
of $\langle\beta_{T}\rangle$$\simeq$0.55 at 130 GeV, and $\sim$0.60 at
200 GeV. These flow velocities appear higher than the values
of $\sim$0.50 obtained at the SPS~\cite{SPS_blast}, indicating
stronger flow. However, some deviation from
the global blast wave fit is seen in
the multistrange baryons at RHIC~\cite{Christophe}.
One additional note is that the fit kinetic
freezeout temperature parameters show at most a mild decrease
going from SPS ($T_{\rm fo}$$\simeq$110 MeV) to RHIC
($T_{\rm fo}$$\simeq$100 MeV) energies.

\section{In-Medium Decay}

One exciting prospect is the possibility to probe the time scale of
hadronic rescattering via in-medium decay. The principle here
is that the daughters of a particle which decays during the
rescattering process have the opportunity to rescatter. With
altered momenta in either or both of the decay daughters,
reconstruction of the parent particle through invariant mass
of the daughters is no longer possible. Particles whose
decay time scales are on the order of the lifetime of the rescattering
period
(perhaps a few fm/$c$) might then have a suppression in their
reconstructed yield proportional to that lifetime.
One must, however, make some assumption as to the
expected yield in order to determine the fraction lost.
There also remains an argument that some additional production
of the parent resonances might occur during rescattering.
This would dilute the measurement of the fraction lost.

\begin{figure}[t]
\begin{center}
\vspace{-0.7cm}
\hspace{-0.3cm}
\includegraphics[width=0.51 \textwidth]{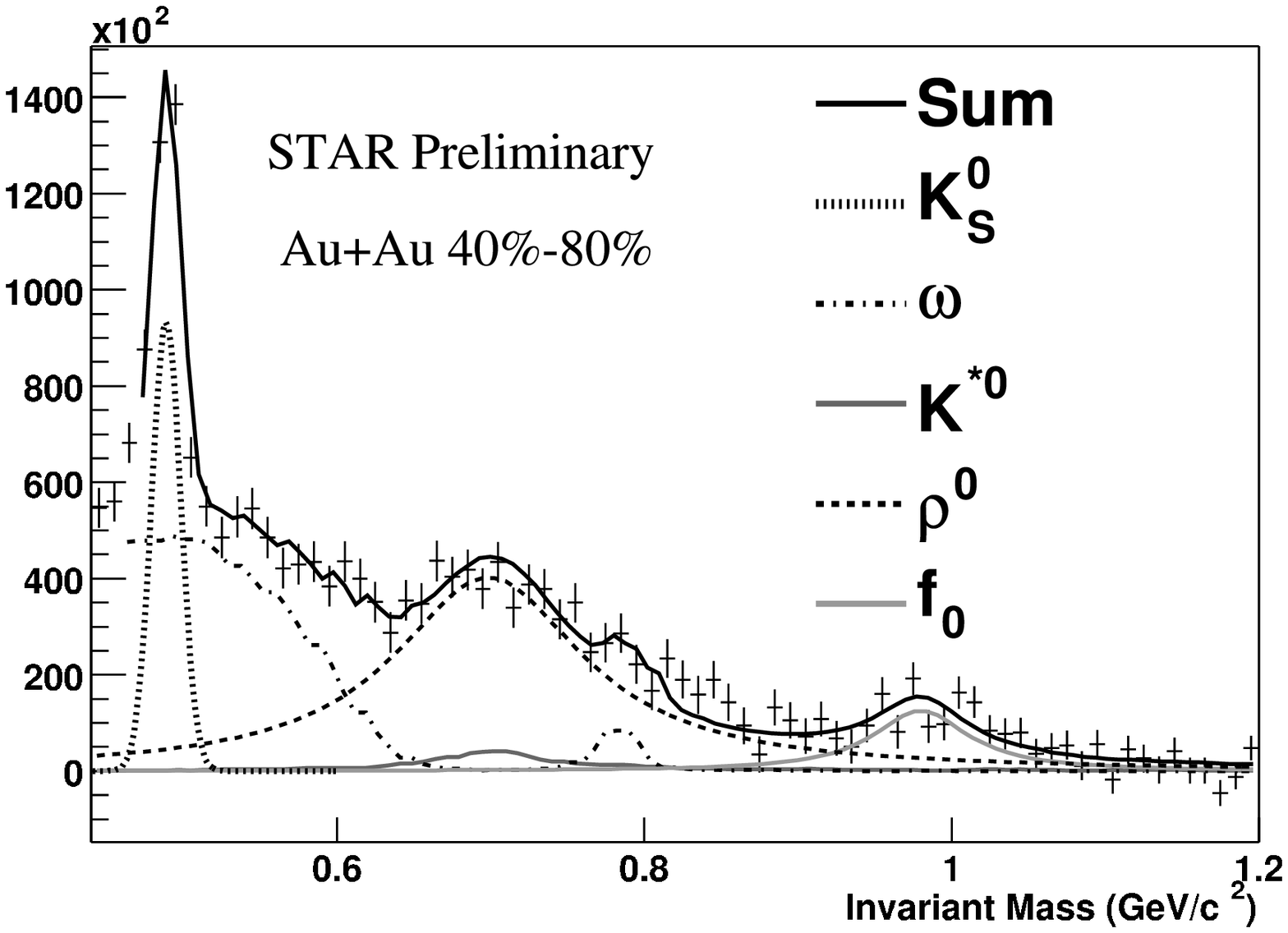}
\hspace{-0.6cm}
\includegraphics[width=0.51 \textwidth]{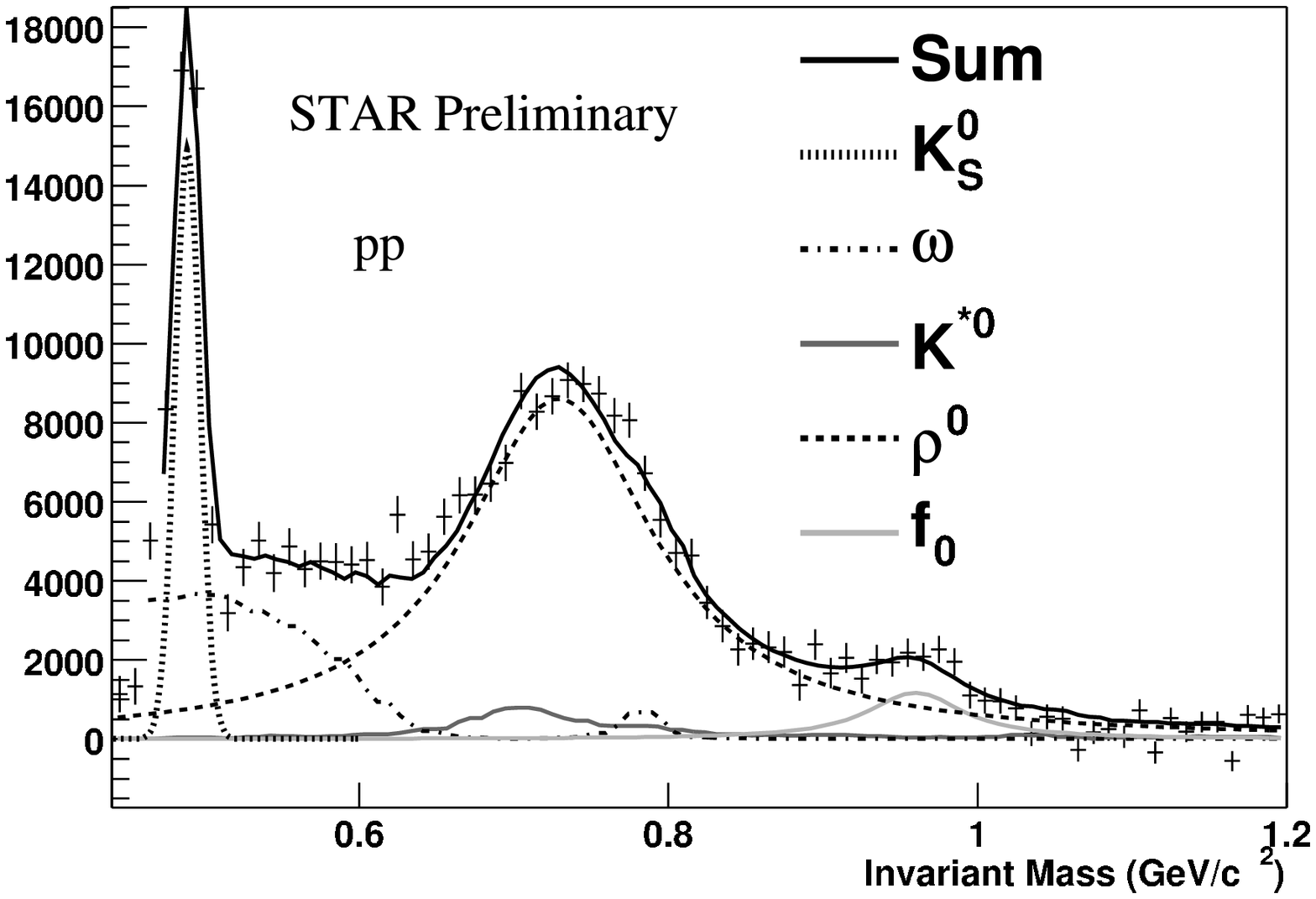}
\vspace{-1.4cm}
\end{center}
\caption{\footnotesize The $\pi^+\pi^-$ invariant mass distributions after background
subtraction for the 40\%-80\% of the hadronic \AuAu~cross section
(left) and for \pp~interactions (right) at \snn{200}.}
\vspace{-0.5cm}
\label{fi:decays}
\end{figure}

To gain some further understanding, STAR has measured
several short-lived species, and has more to come. As an
example, Fig.~\ref{fi:decays} shows the $\pi^+\pi^-$
invariant mass spectra from \AuAu~and \pp~data at \snn{200}.
Here we see evidence for such short-lived resonances as
the $K^{*0}$(892), $\rho$(770), and f$_0$(980). Work is also
in progress to measure the $\eta^0$, $K^{*+,-}$, $\Delta$, $\Lambda$(1520), and
$\Sigma$(1385) (for the first time in heavy ion collisions),
along with the $\phi$, which was presented
at this conference~\cite{Geno}.

\begin{wrapfigure}{r}{0.50 \textwidth}
\begin{center}
\vspace{-1.2cm}
\includegraphics[width=0.50 \textwidth]{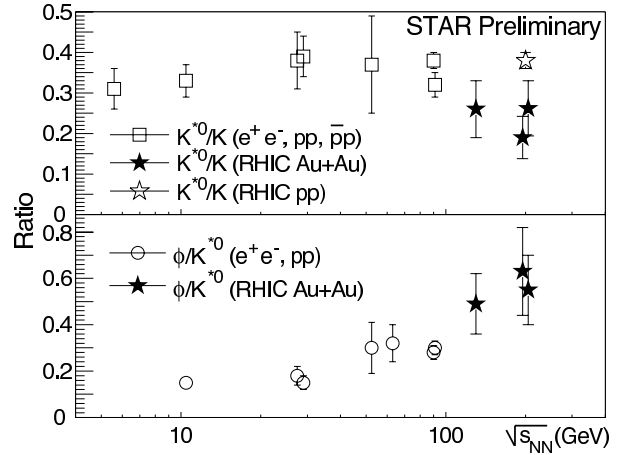}
\vspace{-1.2cm}
\end{center}
\caption{\footnotesize $K^{*0}$/$K$ and $\phi$/$K^{*0}$ yield ratios as a function
of colliding system and energy. For more details, see Ref.~\cite{Patricia}.}
\label{fi:kstar}
\vspace{-1.4cm}
\end{wrapfigure}

In particular, the yield of the $K^{*0}$ has been obtained~\cite{Patricia}
and was already shown in Fig.~\ref{fi:thermal_fit} to not differ notably
from a thermal model for central \AuAu~collisions.
A more detailed comparison can be made by
taking the ratio of the $K^{*0}$ yield to that of the $K$ and $\phi$.
These ratios are shown in Fig.~\ref{fi:kstar} along with the same ratios
for other colliding systems and energies. On this level, it appears
that $K^{*0}$/$K$ is to some degree suppressed
in heavy ion collisions at
RHIC energies, while the $\phi$/$K^{*0}$ ratio appears high. Together
these hint at a slightly low value for the yield of the $K^{*0}$.
This could be
construed as loss due to in-medium decay. However, with a
$c\tau$ shorter than the size of a $Au$ nucleus (and average
velocities considerably slower than $c$), such losses
are actually rather small. This is consistent with either a very short
lifetime of the rescattering period~\cite{rescatter}, or a loss rate only
slightly higher than the regeneration rate during rescattering.

This is only one in a series of observables which may
help us understand the time scales involved in the collision
evolution. STAR is actively working on measurements via
other techniques such as HBT correlations and balance functions
to provide more handles on time and size scales. Several results
from these studies have also been presented at this conference~\cite{HBT}.

\section{Photonic $\pi^0$ Decay}

A notable demonstration of STAR's capabilities comes in measuring
the spectra of $\pi^0$ through the 2$\gamma$ decay
channel, via the subsequent conversion of each $\gamma$
into an $e^+e^-$ pair in the structural material of the
detectors. Details of the analysis for this 4-body
final state can be found in Ref.~\cite{Ian}.

\begin{wrapfigure}{r}{0.41 \textwidth}
\begin{center}
\vspace{-1.3cm}
\includegraphics[width=0.43 \textwidth]{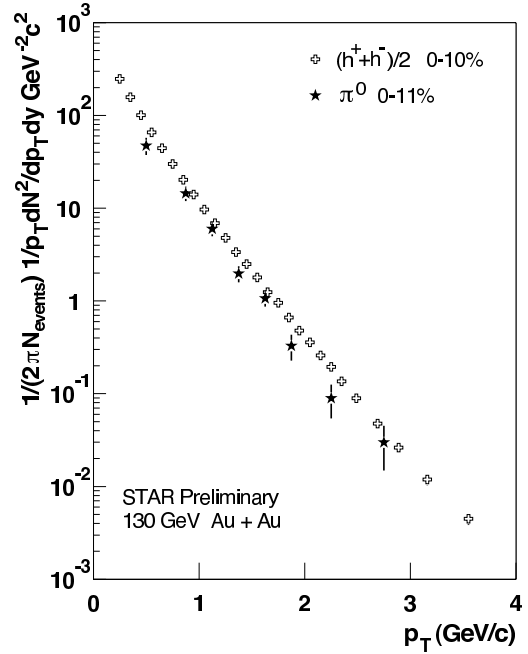}
\vspace{-1.2cm}
\end{center}
\caption{\footnotesize Spectra of $\pi^0$ and h$^{+,-}$
from central \AuAu~collisions at \snn{130}
in STAR. Errors shown on $\pi^0$ data are statistical and do not
include an overall normalization systematic
uncertainty of $\pm$49\%.
For $\pi^0$ spectra from other centralities
and further details, see Ref.~\cite{Ian}.}
\label{fi:pi0_spectra}
\vspace{-1.7 cm}
\end{wrapfigure}

Spectra for the $\pi^0$ are measured for several centralities.
As seen in Fig.~\ref{fi:pi0_spectra}, the shape
compares well
with the soft negative and positive hadron spectra for
central \AuAu~collisions at 130 GeV (which
are dominated by $\pi^-$ and $\pi^+$ respectively at low \pt).
An overall normalization uncertainty of $\pm$49\% not
indicated in the shown errors remains.
It will be exciting to see what can be done with the improvements
available for making
this measurement in the 200 GeV data
and beyond from STAR. One addressable topic may be
determination of the $\pi^0$$\rightarrow$$\gamma\gamma$
contributions to the photon spectra. This and the
$\eta^0$$\rightarrow$$\gamma\gamma$ decay alone
account for $>$97\% of the inclusive photon spectra
in \PbPb~collisions at \snn{17}~\cite{sps_photons}.
STAR measures the inclusive $\gamma$ spectra (currently via
the method outlined above using the TPC, and ahead
with the electromagnetic calorimeter), and has observed an $\eta^0$
signal~\cite{Ian};  the future looks intriguing.

\section{Summary}

Data from the STAR Experiment in the soft regime at midrapidity
probe bulk properties of the evolution of heavy ion collisions at
RHIC energies.
Net baryons from these collisions continue to decrease
with increasing energy, but are still non-zero. This means that
pair production processes behind baryon yields are now dominant,
but do not account for all baryon number; there must remain some
transport of baryon number even over $\sim$5 and $\sim$6 units of rapidity
at 130 and 200 GeV respectively.
Also, the \mpt~systematics of inclusive hadron yields provide
additional information about initial conditions for particle production.
Further investigations into photon spectra may
help us understand the early environment of the collision system.

Particle yields seem, for the most part, to be well-described phenomenologically
by statistical models. The data support thermal, chemical equilibrium
fits to the yields. Such fits proffer a system with thermodynamic properties
of temperature and baryo-chemical potential in the region of the phase
change predicted by lattice QCD calculations.

Evidence for transverse radial flow is observed in the transverse
momentum spectra of observed particles. This is consistent
with the observation of strong elliptic flow by STAR~\cite{flow}.
In addition, apparently slightly suppressed $K^{*0}$
yields hint at hadronic rescattering which is short in
duration if regeneration of the $K^{*0}$ resonance
is much less
than the losses caused by rescattering of in-medium decay daughters.

The STAR Experiment at RHIC continues to address a wide variety
of physics topics in heavy ion and nucleon collisions. Topics explored
through measurements in the soft physics regime are strengths
of the experiment and will persist in providing a better understanding
of the collision dynamics throughout the collision evolution.

\end{document}